\documentclass[useAMS,usegraphicx]{mn2e}
\usepackage{times}

\usepackage{multirow}
\usepackage{epsfig}
\usepackage{subfigure}

\title[MCG--03-34-63 ULX~1] {Have we detected the most
  luminous ULX so far?}  \author[G.\ Miniutti et al.]  {G.\
  Miniutti$^1$\thanks{miniutti@ast.cam.ac.uk}, G.\ Ponti$^{2,3}$, M.\
  Dadina$^{3}$, M.\ Cappi$^{3}$, G.\ Malaguti$^{3}$, A.C.\
  Fabian$^1$ and \and P.\ Gandhi$^{1,4}$\\
  $^1$Institute of Astronomy, Madingley Road, Cambridge CB3 0HA \\
  $^2$ Dipartimento di Astronomia dell' Universit\'a degli Studi di
  Bologna, via Ranzani 1, I--40127, Bologna, Italy\\
  $^3$ IASF/INAF Bologna, via Gobetti 101, I--40129,
  Bologna, Italy \\
  $^4$ European Southern Observatory, Alonso de Cordova 3107, Casilla
  19001, Santiago, Chile }

\pagerange{\pageref{firstpage}--\pageref{lastpage}}
\pubyear{2001}

\usepackage{times}

\begin{document}

\label{firstpage}

 \maketitle

 \begin{abstract} 
%{\bf{
   We report the {\it XMM--Newton} detection of a moderately bright
   X--ray source (F$_{0.5-7}\sim 8.2\times
   10^{-14}$~erg~cm$^{-2}$~s$^{-1}$) superimposed on the outer arms of
   the inactive spiral galaxy MCG--03-34-63 (z=0.0213).  It is clearly
   offset from the nucleus (by about 19'') but well within the
   D$_{25}$ ellipse of the galaxy, just along its bar axis. The field
   has also been observed with the {\it Hubble Space Telescope} (HST)
   enabling us to compute a lower limit of $> 94$ on the X--ray to
   optical flux ratio which, together with the X--ray spectrum of the
   source, argues against a background AGN. On the other hand, the
   detection of excess X--ray absorption and the lack of a bright
   optical counterpart argue against foreground contamination.
   Short--timescale variability is observed, ruling out the hypothesis
   of a particularly powerful supernova.  If it is associated with the
   apparent host galaxy, the source is the most powerful
   Ultra--Luminous X--ray source (ULX) detected so far with a peak
   luminosity of $\sim 1.35\times 10^{41}$~erg~s$^{-1}$ in the
   0.5--7~keV band.  If confirmed by future multi--wavelength
   observations, the inferred bolometric luminosity ($\sim 3\times
   10^{41}$~erg~s$^{-1}$) requires a rather extreme beaming factor
   (larger than 115) to accommodate accretion onto a stellar--mass
   black hole of $20~M_\odot$ and the source could represent instead
   one of the best intermediate--mass black hole candidate so far. If
   beaming is excluded, the Eddington limit implies a mass of
   $>2300~M_\odot$ for the accreting compact object.
\end{abstract}


\section{Introduction}

All known black holes belong to two families: stellar--mass black
holes are seen in X--ray binaries, while super--massive ones are
present in the centres of galaxy bulges sometimes revealing themselves
as Active Galactic Nuclei (AGN). While the former have masses up to
$\sim 20 M_\odot$ (e.g. Fryer \& Kalogera 2001), the latter have
masses in the range $\sim 10^6$--$10^9~M_\odot$, the smaller--mass
super--massive black hole to date being that in NGC~4395 with a mass
of a few times $10^{5}~M_\odot$ (Peterson et al. 2005). Although it
has long been thought that intermediate--mass black holes (IMBH) with
masses $\sim 10^2$--$10^4~M_\odot$ may form in dense stellar clusters
(e.g. Frank \& Rees 1976; Portegies Zwart et al. 1999), there are no
known IMBHs filling the mass--gap between the two known families. If
present, active IMBHs may reveal themselves as accreting X--ray
sources exceeding by a large factor the Eddington luminosity of
stellar--mass black holes (L$_{\rm{20~M_\odot}}^{\rm{Edd}} = 2.6
\times 10^{39}$~erg~s$^{-1}$ for a ``maximal--mass'' stellar--mass
black hole of $20~M_\odot$).

\begin{figure*}
\begin{center}
\hbox{
{\hspace{0.3cm}}
\epsfig{file=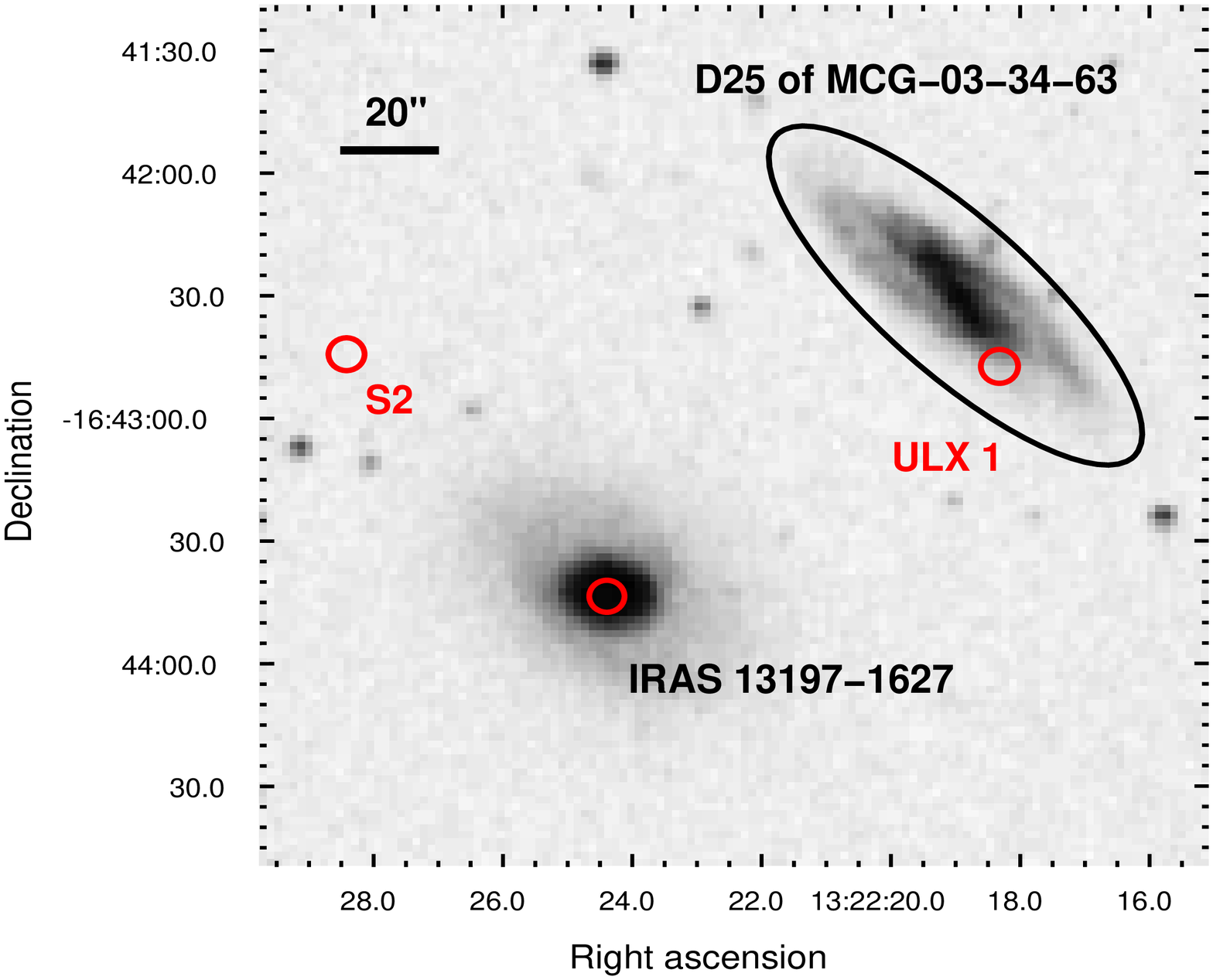,width=0.43\textwidth,height=0.38\textwidth}
{\hspace{2.0cm}}
\epsfig{file=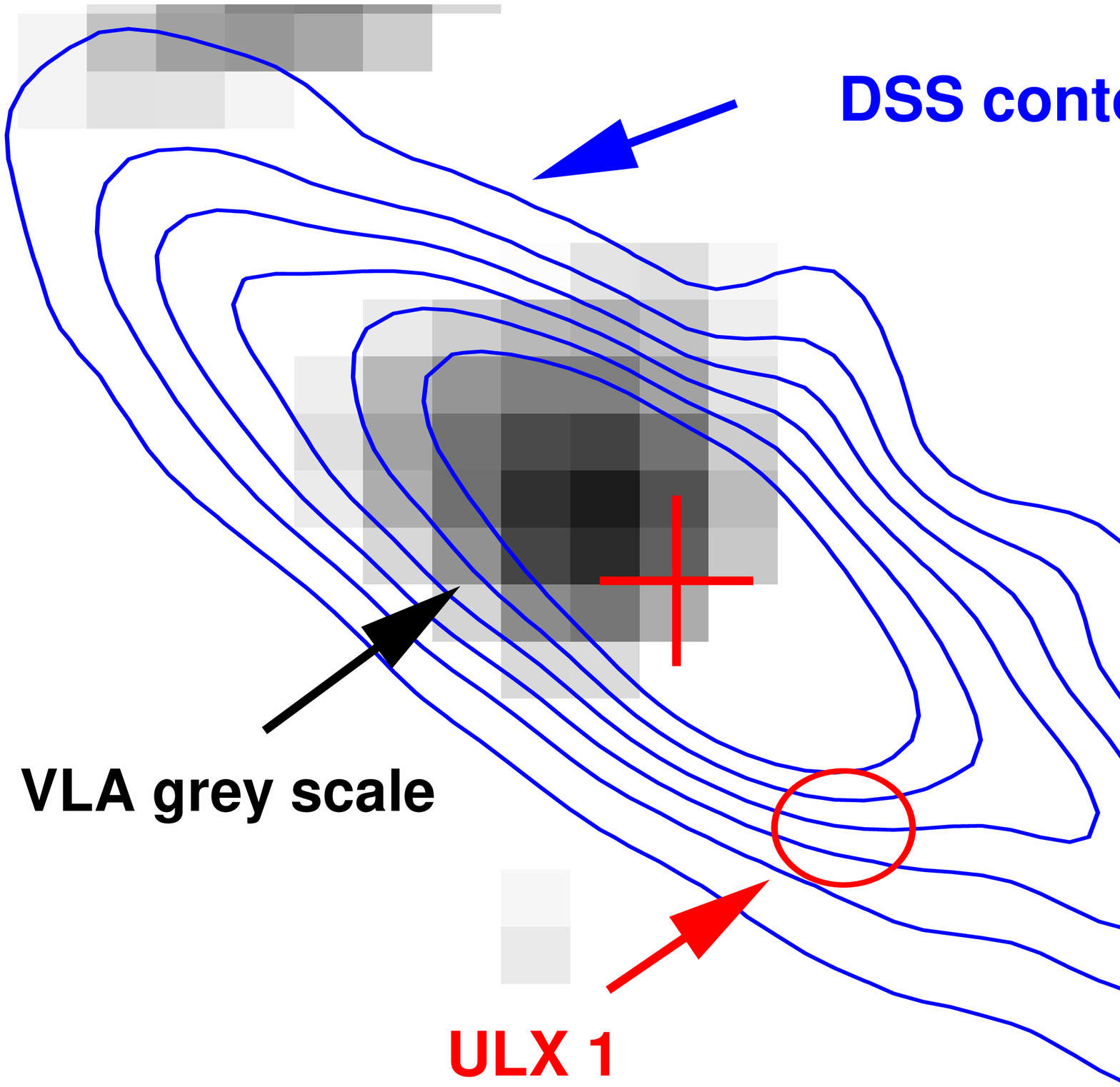,width=0.3\textwidth,height=0.3\textwidth}
}
{\vspace{0.5cm}}
\hbox{
{\hspace{0.9cm}}
%\epsfig{file=hst1paper.eps,width=0.42\textwidth,height=0.29\textwidth}
%{\hspace{1.0cm}}
%\epsfig{file=hst2paper.eps,width=0.39\textwidth,height=0.29\textwidth}
%}
\epsfig{file=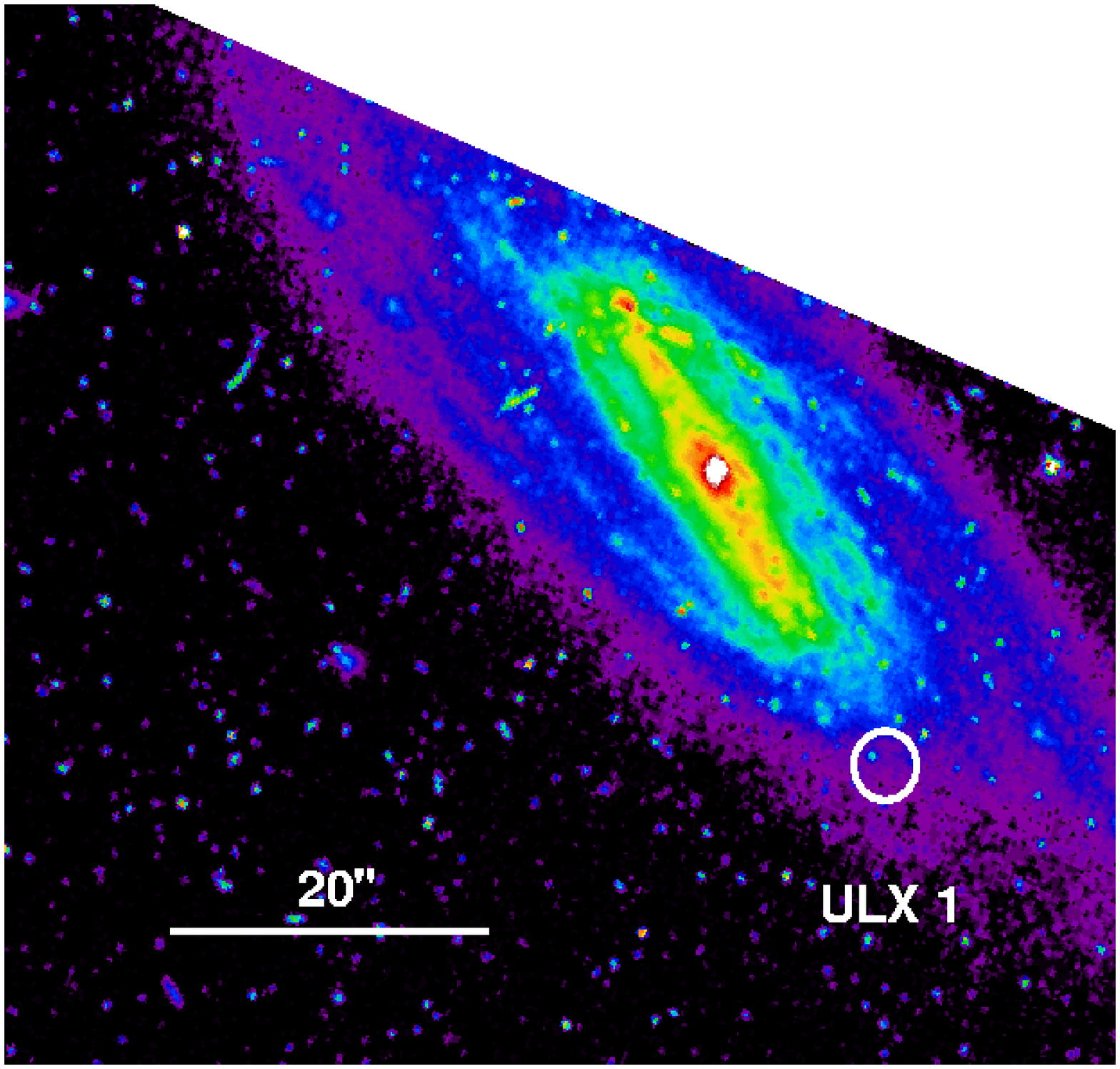,width=0.4\textwidth,height=0.29\textwidth}
{\hspace{2.2cm}}
\epsfig{file=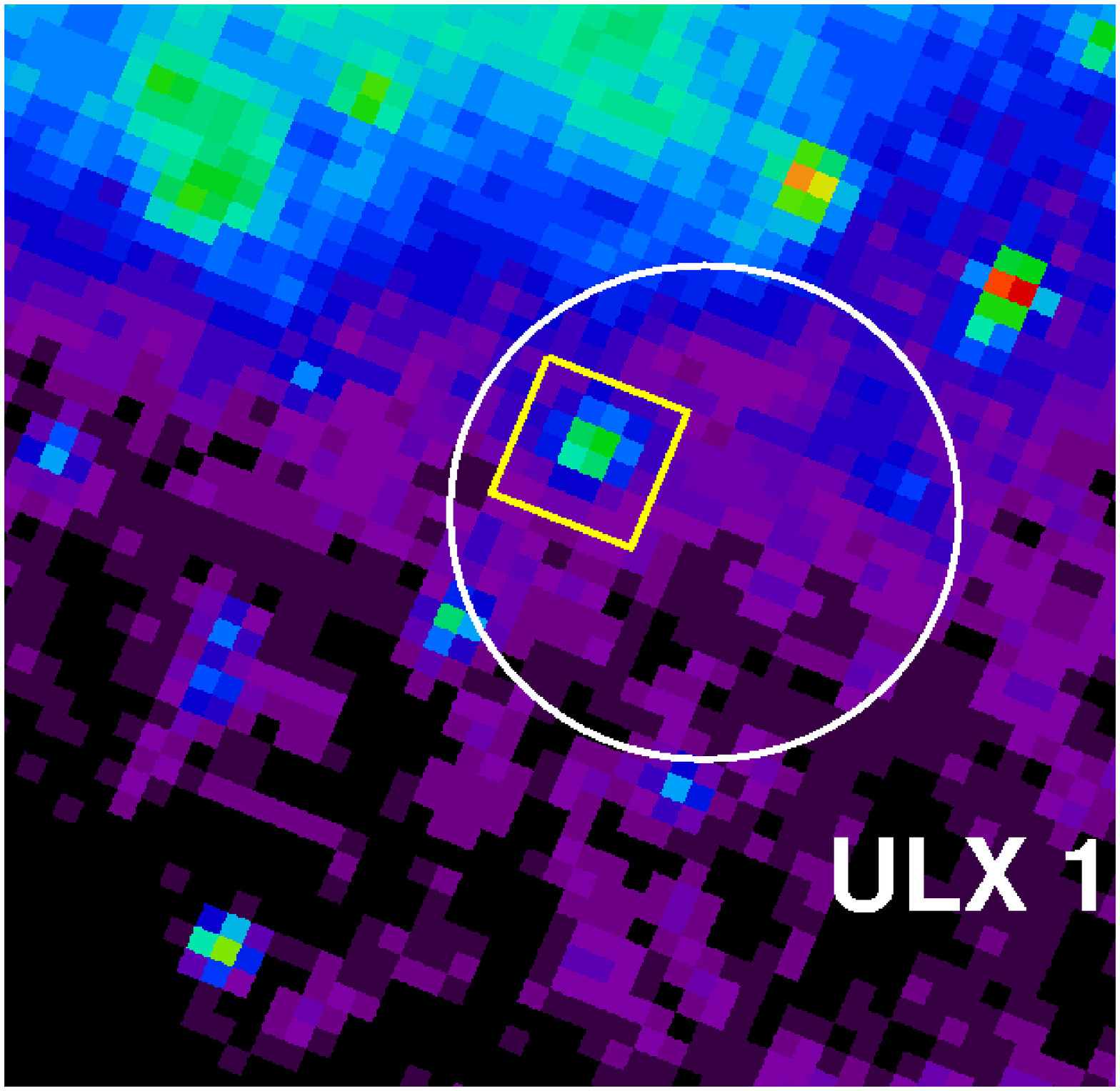,width=0.33\textwidth,height=0.29\textwidth}
}
{\vspace{-0.2cm}}
\end{center}
\caption{For all images, North is upwards and East to the Left. {\bf
    Top Left:} a 3.5'$\times$3.5' DSS optical image in the region of
  IRAS~13197--1627 and MCG--03-34-63 is shown together with the
  detected X--ray sources (red circles). We also show the D$_{25}$
  ellipse of MCG--03-34-63 as a black ellipse. {\bf Top Right:} the
  grey scale is the VLA~21~cm image in the region of of the galaxy
  MCG--03-34-63. Optical contours from the DSS are shown in blue.  The
  red circle is centred on the {\it XMM--Newton} position of ULX~1 and
  the red cross is the optical centre of MCG--03-34-63. {\bf Bottom
    Left:} a portion of the HST WFPC2 500s exposure (F606W filter) of
  MCG--03-34-63 is shown together with the {\it XMM--Newton}
  error--circle (2'' in radius) centred on ULX~1.  {\bf Bottom
    Right:} expanded view of the HST image around the X--ray
  error--circle. The yellow box is centred on the brightest optical
  source within the {\it XMM--Newton} error--circle and represents the
  position of the most likely optical counterpart of ULX~1.}
\end{figure*}

Ultra--Luminous X--ray sources (ULX) are off--nuclear point--like
X--ray sources seen in other galaxies (than the Milky Way) with
luminosities exceeding L$_{\rm{20~M_\odot}}^{\rm{Edd}}$ (see e.g.
Colbert \& Mushotzky 1999; Mushotzky 2004). Since the Eddington
argument implies a lower limit of $20~M_\odot$ on the mass of the
central object, ULXs are often regarded as IMBH--candidates (see e.g.
Miller \& Colbert 2003; Fabbiano 2005). However, inferring a lower
limit on the mass of an accreting compact object only from its
bolometric luminosity can lead to misleading results. This is because,
if potential anisotropies of emission (e.g. beaming, see Reynolds et
al. 1997; King et al. 2001) or accretion (e.g. radiation--driven
inhomogeneous accretion, see Begelman 2002) are not taken into
account, the lower limit on the mass of the object may be severely
over--estimated. Both the beaming and inhomogeneous accretion
scenarios can be invoked to explain luminosities up to a few times
$10^{40}$~erg~s$^{-1}$ with accretion on standard stellar--mass black
holes. However, geometric beaming (e.g. a funnel geometry) and inhomogeneous
accretion can only provide an effective luminosity exceeding the
Eddington limit by a factor $\sim$23 (Madau 1988) and $\sim$10,
(Ruszkowski \& Begelman 2003) respectively. Thus, for ULXs with
luminosities exceeding $10^{41}$~erg~s$^{-1}$, relativistic beaming
seem the only option to avoid the presence of an IMBH. It is thus
clear that the best IMBH--candidates are ULXs in the high end of the
luminosity function with luminosity approaching (or possibly above)
$10^{41}$~erg~s$^{-1}$. The best IMBH--candidate so far is a ULX in
the galaxy M~82. Its extreme luminosity (up to $\sim 9\times
10^{40}$~erg~s$^{-1}$), together with the detection of a 54~mHz QPO
and an Fe K$\alpha$ line (which both argue against a beamed source),
make it difficult not to invoke an IMBH (Matsumoto et al.  2001;
Kaaret et al. 2001; Strohmayer \& Mushotzky 2003).

Here we report the {\it XMM--Newton} detection of an X--ray source
which is plausibly a ULX in the galaxy MCG--03-34-63. If it is at the
distance of its apparent host galaxy, the bolometric luminosity of $\sim 3
\times 10^{41}$~erg~s$^{-1}$, exceeds by a factor $> 115$ the
Eddington luminosity for a $20~M_\odot$ stellar--mass black hole,
making it the most luminous ULX ever detected, and possibly one of the
best IMBH--candidate so far.

\section{MCG--03-34-63 ULX--1}

{\it XMM--Newton} observed the active galaxy IRAS~13197--1627
(z=0.0165) on 2005 January 24 for about 45~ks (Miniutti et al.  2006).
IRAS~13197--1627 is not isolated in the sky and the inactive spiral
galaxy MCG--03-34-63 (z=0.0213) is only 1.8' away towards NW, in the
field of view of all the {\it XMM--Newton} EPIC cameras. MCG--03-34-63
has been often confused with IRAS~13197--1627 (a.k.a. MCG--03-34-64)
until 1996, where it was detected in the radio with a total flux of
93.4~mJy at 4.9~GHz (Colbert et al. 1996). It is also a relatively
luminous IR galaxy with $L_{40-122\mu m} = 2.1\times 10^{10}~L_\odot$,
most likely associated with star--formation (Surace, Sanders \& Mazzarella
2004).

In the top left panel of Fig.~1, we show the optical Digitized Sky
Survey (DSS) image centred between the two galaxies. We also show as
circles the X--ray sources detected in the field, and the D$_{25}$
ellipse of MCG--03-34-63 (the dimension defined to be equal to
25~mag~arcsec$^{-2}$). Besides IRAS~13197--1627, the second strongest
X--ray source is superimposed on the outer arms of the galaxy
MCG--03-34-63\footnote{A second source (S2 in Fig.~1) is also
  detected. Its X--ray spectrum suggests it is a Compton--thin
  Seyfert~2 galaxy at z$\sim$0.16 (Miniutti et al. in preparation).}.
Its X--ray position is displaced by about 19'' in the SW direction
with respect to the optical centre of the galaxy and its coordinates
(J2000) are RA 13h22m18.3s and DEC -16d42m47.7s. Its official name is
thus XMMU~J132218.3--164247 and we shall call it ULX~1 here for
brevity. The X--ray position of ULX~1 has been determined by running
the source detection task {\tt edetect\_chain} and is corrected by
cross--correlating 4 X--ray sources with the USNO--B1.0 optical
catalogue (Monet et al.  2003).  The 1$\sigma$ r.m.s.  absolute
astrometric accuracy of {\it XMM--Newton}\footnote{see Kirsh et al.
  2006, {\it XMM--Newton} calibration document XMM-SOC-CAL-TN-0018 at
  {\tt http://xmm.vilspa.esa.es}} is about 2''. Here, we make a
conservative choice and we assume an error--circle of 2'' in radius.
The X--ray position will be soon known with much better accuracy
thanks to an approved {\it Chandra} observation. No optical
point--like counterpart is seen in the DSS meaning that the source is
fainter than $\sim$21 in R and $\sim$22.5 in B.

 We also analysed the Optical Monitor (OM) mosaic of the field images
 (UVW1 filter).  The UV contours trace MCG--03-34-63, peak at about
 17'' NE of the galaxy centre, and no point--like source corresponds
 to ULX~1. Radio images (VLA 21~cm, 1993 May) have been analysed as
 well and, besides the AGN IRAS~13197--1627, there is an excess within the
 D$_{25}$ ellipse of MCG--03-34-63 which is however not consistent
 with ULX~1 nor exactly centred on the galaxy optical centre (see top
 right panel of Fig.~1).  At 4.9~GHz, the nucleus of MCG--03-34-63 is
 clearly detected (Colbert et al. 1996), but there is no radio excess
 at the position of ULX~1.

The Hubble Space Telescope (HST) also observed the region in July 1994
for 500~s with the WFPC2 and the F606W filter applied. A portion of
the HST image is shown in the bottom left panel of Fig.~1 together
with a white circle of 2'' radius centred on the X--ray position of
ULX~1 and representing the {\it XMM--Newton} error--circle. The HST
astrometry has been corrected by using 5 common sources with the
USNO--B1.0 catalogue. It is interesting to notice that ULX~1 lies
along the bar axis of MCG--03-34-63 where the interstellar medium may
be stressed and where off--nuclear X--ray sources have been found in
other galaxies (see e.g. in NGC~1672, Brandt et al. 1996). In the
bottom right panel of Fig.~1, we show a zoom into the HST image. The
brightest optical source within the {\it XMM--Newton} error--circle is
marked with a yellow box and represents the most likely optical
counterpart (if any) of ULX~1. Another faint source is present in the
error--circle and two are seen at its edges towards SE.

In the X--rays, the field has been observed with {\it ASCA} and {\it
  BeppoSAX}, but the moderate angular resolution of the two X--ray
mission detectors does not allow us to distinguish between the bright
AGN IRAS~13197--1627 and ULX~1. IRAS~13197--1627 is detected by {\it
  Swift/BAT} (Markwardt et al. 2005).  However, given the 17' angular
resolution nothing can be said on ULX~1. No pointed {\it ROSAT}
observation exists and ULX~1 is too faint to be detected in the ROSAT
All Sky Survey.

%The
%brightest HST source within 4'' of ULX~1 is seen 3'' NW of the ULX~1
%position. Although it is ouside the {\it XMM--Newton} error--circle,
%we report here its magnitude (m$_{\rm{F606W}}=23.9$, $R\sim23.2$) for completeness.

%Several point--sources are present in the {\it XMM--Newton}
%error--box, including one at the centre of the X--ray error--box. As detailed
%later, the central most likely optical counterpart has a magnitude of
%M$_{\rm{F606W}}=24.5$ in the F606W HST filter, while the brightest HST
%source, 4'' NW from the centre and superimposed on the edge of the
%{\it XMM--Newton} error--box, has M$_{\rm{F606W}}=23.9$.

\section{X--ray data analysis}

ULX~1 provides 457 background--corrected counts in the EPIC--pn
camera, and about 300 in each of the MOS detectors in the 0.5--7~keV band. We
extracted  the X--ray light curve of ULX~1 from the MOS+pn
detectors to search for short--timescale  variability. Given the
low count rate (about 0.025~Cts/s), we chose a relatively long time
bin of 2.9~ks in order to reduce the errors on the individual data
points. The resulting background-subtracted light curve is shown in
Fig.~2. A fit with a constant gives $\chi^2=27.3$ for 12 dof, showing
the the source is variable at the 99.3 per cent confidence level. By
applying the Kolmogorov--Smirnov test, the significance of the
variability is increased to the 99.94 per cent level.

\begin{figure}
 \begin{center}
 \includegraphics[width=0.35\textwidth,height=0.455\textwidth,angle=-90]{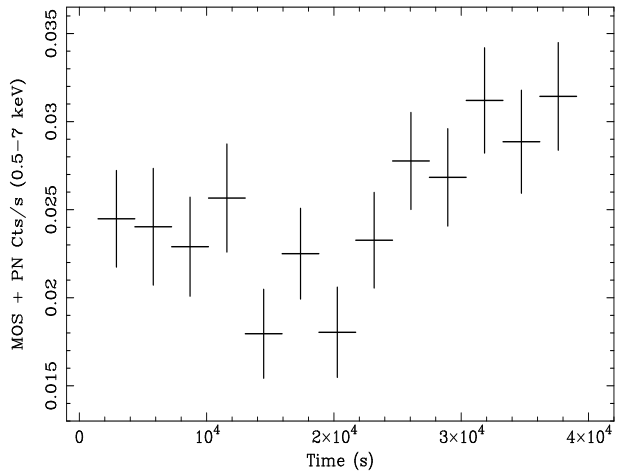}
{\vspace{-0.2cm}}
\end{center}
\caption{We show the background subtracted light curve of ULX~1
  obtained from the MOS+pn data in the 0.5--7~keV band (binsize of
  2.9~ks). The Kolmogorov--Smirnov probability that the light curve is
  constant in $5.7 \times 10^{-4}$ only.}
 \end{figure}

\begin{figure}
\begin{center}
\includegraphics[width=0.35\textwidth,height=0.455\textwidth,angle=-90]{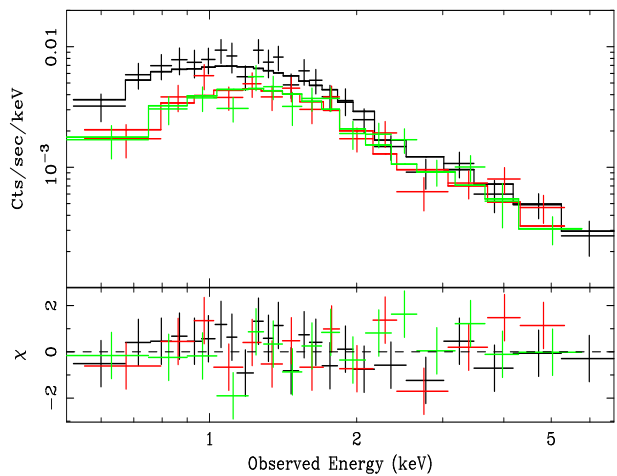}
{\vspace{-0.2cm}}
\end{center}
\caption{The pn (black) and MOS (red and green) spectrum of ULX~1 is
  shown together with the best--fitting absorbed power law model and
  the resulting residuals (see text for discussion).}
\end{figure}

We also extracted the X--ray spectra from both the pn and MOS cameras
in the same 0.5--7~keV band. As a first attempt to describe the X--ray
spectral shape of ULX~1, we consider an absorbed power law model and
we fix the absorbing column to the Galactic value, i.e.  at $5.8\times
10^{20}$~cm$^{-2}$ (Dickey \& Lockman 1990). The joint fit to the pn
and MOS data is unacceptable ($\chi^2=74$ for 51 dof) because the
spectrum is more curved than the model, suggesting the presence of
excess absorption. We then keep the Galactic column fixed, but add a
second (neutral) absorption component. The statistical quality of the
fit improves dramatically and we obtain $\chi^2=37$ for 50 dof. We
measure an excess absorption column density of $2.3^{+0.6}_{-0.4}
\times 10^{21}$~cm$^{-2}$, while the power law has a slope
$\Gamma=1.95^{+0.18}_{-0.16}$. The spectra, best--fit model and
residuals are shown in Fig.~3. We point out that the observed amount
of excess absorption is typical for ULXs, suggesting that part of it
originates in the ULX environment (Swartz et al. 2004; Winter,
Mushotzky \& Reynolds 2006). The X--ray spectral slope also compares
very well with other ULXs and lies at the boundary between
``low--state'' and ``high--state'' sources (as classified by Winter et
al. 2006).

The observed 0.5--7~keV flux is $8.2\times
10^{-14}$~erg~cm$^{-2}$~s$^{-1}$. By extrapolating the model up to
10~keV, the source has a 0.5--10~keV flux of $9.8\times
10^{-14}$~erg~cm$^{-2}$~s$^{-1}$. Note that the observed variability
(see Fig.~2) implies that during the observation ULX~1 reached a flux
of about $\sim 1.2\times 10^{-13}$~erg~cm$^{-2}$~s$^{-1}$ in the
0.5--10~keV band. 
%As discussed in detail below, if the distance of
%MCG--03-34-63 is assumed, the observed intrinsic luminosity of ULX~1 is
%L$_{0.5-7} = 1.1 \times 10^{41}$~erg~s$^{-1}$ (with $H_0=70$~ks~s$^{-1}$,
%$\Lambda_0=0.73$, and $q_0=0$).

\subsection{The extreme luminosity of ULX~1}

The excess absorption we detect and the lack of a bright optical
counterpart (see discussion below) makes it unlikely that the source
is a foreground object, though we cannot exclude it. If so, the X--ray
luminosity of the source can be computed as a lower limit under the
hypothesis that it is at least at the distance of MCG--03-34-63
(z=0.0213). By using our best--fitting absorbed power--law model, it
turns out that ULX~1 has an intrinsic luminosity L$_{0.5-7} = 1.1
\times 10^{41}$~erg~s$^{-1}$ in the 0.5--7~keV band and L$_{0.5-10} =
1.3 \times 10^{41}$~erg~s$^{-1}$ once extrapolated to 10~keV (with
$H_0=70$~ks~s$^{-1}$, $\Lambda_0=0.73$, and $q_0=0$). If the
light curve variability is taken into account, the source reached a
luminosity of $\sim 1.35 \times 10^{41}$~erg~s$^{-1}$ ($\sim 1.6
\times 10^{41}$~erg~s$^{-1}$) in the 0.5--7~keV (0.5--10~keV) during
the {\it XMM--Newton} observation.

We point out that of the $\sim$230 ULXs detected and catalogued up to
now (e.g. Liu \& Mirabel 2005; Ptak et al. 2006), only a few have
luminosities exceeding $3\times 10^{40}$~erg~s$^{-1}$. Luminosities of
the order of $10^{41}$~erg~s$^{-1}$ or above are extremely unusual, as
also shown by the luminosity cutoff in the cumulative luminosity
functions (Swartz et al. 2004; Gilfanov, Grimm \& Sunyaev 2004).  This
makes ULX~1 a rare, if not unique, source among the other known ULXs.

%41 have a 2--10~keV
%luminosity at least one order of magnitude lower than ULX~1 and only 3
%(in M~82, NGC~3311, and NGC~4565) have a luminosity a factor of a few
%lower than ULX~1. Besides the ULX in M~82, only 1 ULX (in IC~2597) out
%of 229 in the catalogue by Liu \& Mirabel (2005) approaches the
%luminosity of ULX~1. The high luminosity we observe makes ULX~1 a
%rare, if not unique, source among the other known ULXs.

\subsection{Adding a thermal component}

To test the similarity of the X--ray properties of ULX~1 with
stellar--mass accreting black holes, we also tried two different
models: a multi--colour disc (MCD) spectrum ({\tt DISKBB}) plus power
law, and a Comptonization model ({\tt COMPTT}) that describes Compton
up--scattering in a corona of optical depth $\tau$ and electron
temperature $kT_{\rm el}$ from a Wien distribution of soft seed
photons from the disc (with temperature $kT$). It is worth noting here
that the detection of a thermal component could give important (though
not conclusive) insights on the compact object mass since, for
standard accretion disk, $T_{\rm{disc}} \propto M_{\rm{BH}}^{-1/4}$
(objects hosting black-holes of $\sim 10~M_\odot$ have typical
temperatures of 1~keV).  A cool thermal component from the accretion
disc may thus provide some indication for a particularly massive black
hole (see e.g. Miller, Fabian \& Miller 2004).

However, when applied to the pn and MOS data, neither model provides a
statistically significant improvement with respect to the power law
one. We can only measure 90 per cent upper limits for the temperature of
the putative thermal component of $kT<0.70$~keV (MCD model), and
$kT<0.35$~keV ({\tt COMPTT}).  Thus, the data are consistent with
a cool disc, but the thermal component is by no means required by
the data.

\section{An extreme ULX or a background AGN?}

The extreme luminosity estimated above assumes that ULX~1 lies within
its apparent host galaxy MCG--03-34-63.  Luminosities approaching
$\sim 10^{41}$~erg~s$^{-1}$ can also be reached by extremely powerful
supernovae occurring in dense environments, but given that their
emission either fades or remains constant over a $\sim$1~yr timescale
(e.g.  Schlegel 1995), the observed X--ray short--timescale
variability rules out this hypothesis. However, ULX~1 could be a
background X--ray source. In this case, the inferred luminosity is
only a lower limit and a background AGN could well be invoked to
explain the X--ray power of the source.  This hypothesis can only be
tested observationally by obtaining an optical spectrum of the
counterpart (if any), but we can nevertheless provide two arguments in
favour of an association with MCG--03-34-63.

\subsection{The $\log N$--$\log S$ estimator}

The number of expected  X--ray sources within a given area can be estimated by
assuming that our field is characterised by the same $\log N$--$\log
S$ as observed from deep {\it XMM--Newton} exposures of the Lockman
Hole (Hasinger et al. 2001). Our ULX candidate has a 0.5--2~keV flux
of $2.8\times 10^{-14}$~erg~cm$^{-2}$~s$^{-1}$ and the Lockman Hole
$\log N$--$\log S$ implies the presence of about 30 sources per square
degree with the same or higher flux (we point out that 30 sources is a
conservative estimate). Since the D$_{25}$ ellipse of MCG--03-34-63
covers an area of $2.4 \times 10^{-4}$ square degrees, the number of
expected sources within D$_{25}$ is 0.007. In the 2--10~keV
band, we measure a flux of $7\times 10^{-14}$~erg~cm$^{-2}$~s$^{-1}$
which again corresponds to about 30 sources per square degree
providing the same small number of expected sources.

%{\bf{
As mentioned, this is only the number of expected sources within a
given area (the D$_{25}$ ellipse of MCG--03-34-63) and not directly
the probability that ULX~1 is a background AGN. We provide the above
estimate for comparison with previous works on other ULXs, but we must
point out that an accurate estimate of the true probability that an
X--ray source within a galaxy is a background AGN has yet to be
produced, and is well beyond the purpose of this Letter.
%}}

\subsection{The X--ray to optical flux ratio}

Another possible way to constrain the nature of the source is by
considering the X--ray to optical flux ratio (X/O$\equiv F_X /F_{\rm
  Opt}$). AGN have typical X/O below 10 and detailed studies of the
optically faint sources in the {\it XMM--Newton} HELLAS2XMM and {\it
  Chandra} deep fields surveys reveal that less than 5 per cent of the
hard X--ray selected sources have X/O higher than 90 (Mignoli et al.
2004; Civano, Comastri \& Brusa 2005).

%{\bf{
We have computed the X/O for the brightest, and most likely optical
counterpart, HST source in the {\it XMM--Newton} error--circle (i.e.
the source within the yellow box in the bottom right panel of Fig.~1).
From the HST image, we estimate a magnitude m$_{\rm{F606W}}=24.5$
giving rise to R$\sim 23.8$ (see Holtzman et al. 1995 for the
conversion between HST and Johnson magnitudes). By using the
definition $\log {\rm X/O} = \log {\rm F}_X + 0.4~{\rm R} +5.61$, we
then estimate X/O$\sim 94$ which represents a lower limit on the X/O
of ULX~1 (where F$_X$ is the 2--10~keV flux). As mentioned above,
sources with X/O$>90$ represent less than 5 per cent of the hard
X--ray sources population in {\it Chandra} and {\it XMM--Newton} deep
fields.  Moreover, the X--ray and near infrared data of high--X/O
sources in the {\it Chandra} deep fields and the HELLAS2XMM survey
identify most (if not all) sources as heavily absorbed AGN (Civano,
Comastri \& Brusa 2005; Mignoli et al.  2004). In the case of ULX~1,
the steep photon index and moderate absorption are inconsistent with
this scenario. The detection of an unobscured background AGN with
X/O$>94$ would be a very rare event on its own, and its detection
within the D$_{25}$ ellipse of a spiral galaxy seems highly unlikely,
supporting the idea that ULX~1 is not a background AGN but more likely
an X--ray binary. It should also be pointed out that the ULXs that
have been later classified as background AGN had standard rather than
high X/O (Foschini et al.  2002; Masetti et al. 2003).
%}}

\section{Conclusions}

If ULX~1 is indeed associated with its apparent host galaxy
MCG--03-34-63, its luminosity greatly exceeds that of a neutron star,
and we have to consider an accreting black hole. In this case, by
considering as a template stellar--mass accreting black holes in our
Galaxy, the bolometric luminosity is likely to be a few times higher
than the 2--10~keV luminosity. By extending our spectral model, we
estimate an intrinsic luminosity of $\sim 3\times
10^{41}$~erg~s$^{-1}$ in the 0.2--150~keV band, which we take as a
reasonable proxy for the bolometric luminosity of ULX~1. In the
absence of beaming, the Eddington limit can be used to infer the
presence of a $> 2300~M_\odot$ black hole, i.e. ULX~1 would be powered
by an accreting IMBH.

To reduce the mass of the compact object within the acceptable range
of stellar--mass black holes, the intrinsic luminosity has to be
boosted by a factor $>115$, which could be due to beaming.  The
simplest candidate mechanism to produce X--ray beaming in an accreting
source is to consider a thick accretion disc with a smaller optical
depth over a limited range of angles around the rotation axis
producing thereby a central funnel.  Such a geometry has been
investigated by Madau (1998) who found that if maximal beaming is
considered, the effective observed luminosity ($L_{\rm{eff}}$) can
exceed the Eddington limit by a factor of $\sim$23 (see also Misra \&
Sriram 2003 who derived a much smaller factor of $\sim$5). Thus, even
by considering maximal beaming with $L_{\rm{eff}} = 23\times
L_{\rm{Edd}}$, ULX~1 does still require a black hole with mass $>
100~M_\odot$.  The mass of the central black hole could be in
principle reduced to 10--20$M_\odot$ by considering maximal beaming
{\emph{and}} a source which is super--Eddington by a large factor
(5--10), which seems a rather finely--tuned choice of the parameters.
However, Begelman, King \& Pringle (2006) recently pointed out that
such a scenario is possible though it would predict a strong soft
thermalized component (not required by the X--ray spectrum of ULX~1).
Beaming can also occur if the X--ray emission is associated with a
relativistic jet whose axis is aligned with the line of sight, as
proposed by e.g. Reynolds et al.  (1997) for ULXs. Indeed, a source
with the jet axis aligned within a few degrees with our line of sight
and with a sufficiently high Lorentz factor ($\gamma>4.3$) could
account for the observed luminosity with a sub--Eddington jet from a
stellar--mass black hole.

%Although the arguments we presented against a background AGN
%identification seem robust, they are statistical in nature, and
%exceptions on single sources are always possible. All available
%information point towards a luminous X--ray binary in MCG--03-34-63,
%either relativistically beamed or powerd by an accreting IMBH. It is
%however clear that the real nature of ULX~1 will be better assessed
%only by performing detailed multi--wavelength observations in the
%future.

\section*{Acknowledgements}
Based on observations obtained with XMM-Newton, an ESA science mission
with instruments and contributions directly funded by ESA Member
States and NASA. 
%This research has made use of the NASA/IPAC
%Extragalactic Database (NED) which is operated by the Jet Propulsion
%Laboratory, California Institute of Technology, under contract with
%the National Aeronautics and Space Administration. 
This research made
use of observations made with the NASA/ESA Hubble Space Telescope,
obtained from the Data Archive at the Space Telescope Science
Institute, which is operated by the Association of Universities for
Research in Astronomy, Inc., under NASA contract NAS 5-26555. The
Digitized Sky Surveys were produced at the Space Telescope Science
Institute under U.S. Government grant NAG W-2166 based on data
obtained using the Oschin Schmidt Telescope on Palomar Mountain and
the UK Schmidt Telescope. G. Miniutti thanks the PPARC for support. GP, MD, MC
and G. Malaguti thank ASI financial support under contract
I/023/05/0. ACF thanks the Royal Society for support. PG is supported
by an ESO (European Southern Observatory) Fellowship. We would like to
thank the referee for her/his constructive criticisms and suggestions.

\end{document}